\documentclass[acmsmall,nonacm]{acmart}
\AtBeginDocument{%
  }

\setcopyright{none}
\renewcommand\footnotetextcopyrightpermission[1]{}

\usepackage[most]{tcolorbox}
\newtcolorbox{keyfindingbox}[1]{%
  colback=black!6, colframe=black!75,
  boxrule=0pt, leftrule=1.2pt, rightrule=1.2pt,
  arc=3.5pt, boxsep=0pt,
  left=5pt, right=5pt, top=3pt, bottom=3pt,
  before skip=6pt plus 1pt minus 1pt,
  after skip=6pt plus 1pt minus 1pt,
  before upper={\textbf{#1:}\enspace},
}

\begin{document}

\title{Where Did the Repair First Go Wrong? Localizing the Origins of Silent Failures in Agentic Vulnerability Repair}

\author{Wenji Bai}
\email{wenji.bai@tuni.fi}
\orcid{0000-0002-3283-3483}
\affiliation{%
  \institution{Tampere University}
  \city{Tampere}
  \country{Finland}
}

\author{Muhammad Waseem}
\orcid{0000-0001-7488-2577}
\email{muhammad.waseem@tuni.fi}
\affiliation{%
  \institution{Tampere University}
  \city{Tampere}
  \country{Finland}
}

\author{Zeeshan Rasheed}
\orcid{0000-0001-9655-3096}
\email{zeeshan.rasheed@tuni.fi}
\affiliation{%
  \institution{Tampere University}
  \city{Tampere}
  \country{Finland}
}

\author{Jaakko Peltonen}
\orcid{0000-0003-3485-8585}
\email{jaakko.peltonen@tuni.fi}
\affiliation{%
  \institution{Tampere University}
  \city{Tampere}
  \country{Finland}
}

\author{Pekka Abrahamsson}
\orcid{0000-0002-4360-2226}
\email{pekka.abrahamsson@tuni.fi}
\affiliation{%
  \institution{Tampere University}
  \city{Tampere}
  \country{Finland}
}



\begin{abstract}
Localizing where an LLM-based agent first fails to uphold security during a repair can show which stage of its workflow needs an additional safeguard. This is difficult for silent failures, which are patches that pass syntactic and functional checks but still contain a security vulnerability. Because such patches give no observable failure signal, existing failure attribution methods, which rely on observed task failures and labelled failure steps, are less suited to them. We propose Security Awareness Gap Evaluation (SAGE), a trace-based method that combines an assessment of the security reasoning recorded at each turn with the reconstructed code history to identify the earliest turn at which a repair diverges from the task's security intent. We evaluate SAGE on 95 confirmed silent failures drawn from 3,684 repair traces produced by six agent frameworks and six base models on SecurityEval and CVEfixes. SAGE assigned an origin in 93 cases. Most origins were an unaddressed security requirement or an inadequate defence choice, and only five coincided with the code change itself. When the agent introduced the vulnerable code, the origin preceded the write in 14 of 19 cases. Repeated scoring and a second judge reproduced the origin type more consistently than the exact turn, and agreement was lowest for traces that kept only the final file.
\end{abstract}




\maketitle

\section{Introduction}
\label{sec:introduction}

Large language model (LLM)-based agents increasingly approach software repair as an iterative process of planning, tool use, code editing, and validation~\cite{bouzenia2025repairagent,zhang2024autocoderover}. Yet, passing the available tests does not necessarily establish that the relevant security requirements have been met \cite{zhao2026vibe}. Recent work has identified silent failures (SFs): patches that pass the available syntactic and functional checks while retaining or introducing a security vulnerability~\cite{bai2026passingtestshidesvulnerabilities}. Confirming a silent failure shows that the delivered repair remains insecure, but not where the repair process first went wrong. The same insecure patch may stem from a plan that omits the relevant security requirement, a defence that does not close the weakness, or an implementation that applies an appropriate defence incorrectly. Such a flaw may then pass through later review stages unnoticed. Each origin may call for a different remedy, ranging from prompt refinement to an additional verification checkpoint.

\begin{figure}[t]
  \centering
  \includegraphics[width=\textwidth]{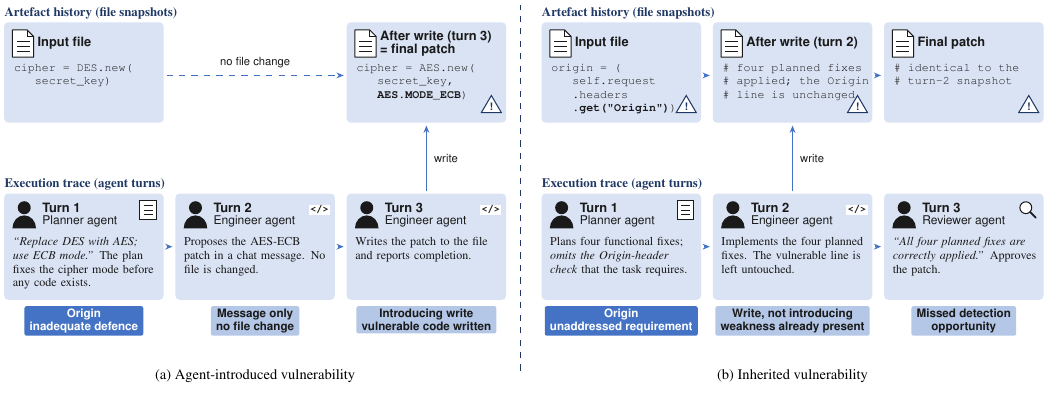}
  \caption{Examples of the two cases in silent failure origin localization.}
  \label{fig:rq1-example}
\end{figure}

We therefore ask where in the execution trace a divergence from the task's security intent first becomes observable. We define this position as the origin of the silent failure: the earliest turn at which the recorded reasoning or artefact history shows such a divergence, as opposed to the write that eventually places the vulnerable code in the file. Figure~\ref{fig:rq1-example} illustrates two ways in which the origin of a silent failure can differ from a code change. In case (a), the planner agent selects AES in ECB mode at turn 1, two turns before the vulnerable code is written. A diff identifies the introducing write, but only the trace reveals the earlier inadequate defence choice. In case (b), the vulnerable code is inherited from the input file and remains untouched, so there is no introducing write to locate. The trace instead records an initial plan that omits the relevant security requirement and a subsequent review that does not revisit the omitted requirement. Final-patch analysis alone is therefore insufficient; origin localization calls for examining the recorded repair reasoning alongside the artefact history, and for separating the origin from both the introducing write and later missed detection opportunities.

Existing research provides useful starting points for analysing agent failures, but does not address the origin of a silent failure in vulnerability repair. Failure taxonomies and attribution methods have mainly been developed and evaluated on executions with an observable task failure~\citep{cemri2026why,zhang2025which,zhang2026agentracer}, whereas a silent failure passes the available checks and its deficiency is established only through separate confirmation. The prior study of silent failures attributes deficiencies to agent roles and workflow stages, but does not provide a systematic rule for localizing the earliest origin turn~\citep{bai2026passingtestshidesvulnerabilities}. 

To address this gap, we propose Security Awareness Gap Evaluation (SAGE), a trace-based method for localizing the origin of a silent failure. SAGE first normalizes framework-specific traces into a common sequence of turns. It then uses an LLM judge, given only the trace prefix available at each turn, to score the security reasoning expressed along four dimensions; every non-zero score must cite a verbatim quotation linked to the raw record (see Section~\ref{sec:sage}). In parallel, it reconstructs the artefact history to determine when the vulnerable code was introduced or left in place. Finally, an explicit decision rule combines the scores and artefact history with the confirmed failure anchor to identify the earliest trace-attributable divergence from the task's security intent (see Section~\ref{sec:localization}). We evaluate the resulting origin assignments through repeated scoring, cross-judge comparison, task-disjoint subsets, and human annotation of artefact-derived outputs. Three research questions (RQs) guide the study:\begin{description}
    \item[\textbf{RQ1.}] Where do silent failures originate, and what trace evidence supports the assigned origins?

    \item[\textbf{RQ2.}] How consistently does trace- and artefact-based evidence identify the same origin across repeated runs, task-disjoint data subsets, and independent judges?

    \item[\textbf{RQ3.}] How does localization reliability vary across silent failure categories, agent frameworks, and base models, and under what conditions does the identified origin become less consistent?
\end{description}

We evaluate SAGE on 95 confirmed silent failures identified among 3,684 valid repair traces generated by six open-source agent frameworks and six base models on tasks from SecurityEval~\cite{siddiq2022securityeval} and CVEfixes~\cite{bhandari2021cvefixes}. SAGE localizes an origin in 97.89\% of cases (93/95). Most identified origins involve unaddressed security requirements or inadequate defence choices. Among the 19 cases in which the agent introduced vulnerable code, the origin precedes the introducing write in 14. Repeated scoring and an independent judge model reproduce the same origin type in most cases. Agreement with human annotators on code provenance and the introducing write is moderate to substantial ($\kappa$ between 0.58 and 0.74) and varies across task subsets. Exact-turn agreement is lowest for the small group of traces that preserve only the final file. 

To summarize, our main contributions are:

\begin{itemize}
    \item \textbf{A trace-based localization method SAGE.} We formulate silent failure localization as identifying the earliest trace-attributable divergence from the task's security intent in an execution that passes functional validation. We propose SAGE, which combines quotation-anchored assessment of expressed security reasoning with reconstructed artefact histories.

    \item \textbf{A multi-source reliability evaluation.} We evaluate SAGE through repeated scoring, cross-judge comparison, task-disjoint subsets, and human annotation.

    \item \textbf{A characterization of localization boundary conditions.} We examine how reliability varies across silent failure categories, agent frameworks, base models, and artefact availability, and identify conditions associated with less consistent localization.
\end{itemize}
\section{Background and Related Work}
\label{sec:related}

\textbf{Security of LLM-generated code and agentic vulnerability repair.} Empirical studies report security weaknesses in LLM-generated code~\citep{pearce2022asleep,asare2023copilot,fu2025security,majdinasab2024copilot,perry2023users} and limitations in LLM vulnerability detection and reasoning~\citep{ding2025primevul,ullah2024cannot}. Datasets such as SecurityEval, LLMSecEval, CVEfixes, and BaxBench support security evaluation through vulnerability prompts, historical fixes, and exploit-based tests~\citep{siddiq2022securityeval,tony2023llmseceval,bhandari2021cvefixes,vero2025baxbench}. Zero-shot and neural repair approaches remain limited on real-world vulnerabilities~\citep{pearce2023examining,wu2023neural}, and attention has moved to agents that locate, edit, and validate fixes over several turns~\citep{bouzenia2025repairagent,zhang2024autocoderover}. Such runs are commonly scored by test passage~\citep{jimenez2024swebench}, yet test-passing patches may still be incorrect~\citep{smith2015cure,wang2026solved,zhou2024patchcorrect}, and a simple non-agentic pipeline can achieve competitive scores~\citep{xia2025agentless}. These studies motivate evaluation beyond final test outcomes, but they do not explain where in an agentic repair trajectory a security failure first becomes attributable.

\textbf{Silent failures in agentic repair.} Bai et al.~\citep{bai2026passingtestshidesvulnerabilities} studied silent failures in agentic vulnerability repair and classified confirmed SF cases by category, agent role, and workflow stage. That work shows that security deficiencies can remain hidden behind patches that pass available syntactic and functional checks. The present study addresses the next diagnostic problem: once a silent failure is confirmed, where in the recorded repair trajectory did the process first diverge from the task's security intent? This requires separating the origin turn from code provenance, the introducing write, and later missed detection opportunities. Prior silent-failure categorization did not provide a systematic method for such turn-level localization or evaluate whether origin assignments are consistent across scoring runs and judges.

\textbf{Failure analysis and attribution in agentic systems.} Taxonomies characterize agent failures by system design, inter-agent misalignment, and task verification~\citep{cemri2026why}, and analyses of repair-agent trajectories associate failed runs with repetitive, non-adaptive action cycles~\citep{bouzenia2025trajectories}. AgentDebug classifies errors by agent module and diagnoses root-cause errors to guide recovery~\citep{zhu2025agentdebug}. Attribution methods identify responsible agents or steps under different evidence conditions. Zhang et al.~\citep{zhang2025which} apply LLM-based attribution to failed multi-agent runs where task failure is observable and step-level labels are available. AgenTracer~\citep{zhang2026agentracer} trains a tracer on failure trajectories constructed through counterfactual replay and fault injection. FAMAS~\citep{ge2026spectrum} ranks agents and actions using spectrum analysis over repeated executions, requiring additional trajectories that may be unavailable in a single-trace audit. DuoTrace~\citep{zhang2026duotrace} learns regular execution patterns from normal trajectories and filters anomalous steps before downstream attribution, but its anomaly scores are not designed to link a step to a confirmed security deficiency in a functionally passing patch. Benchmark studies further show that attribution accuracy depends on the completeness of the supplied evidence~\citep{chen2026elephant} and remains limited on lengthy traces~\citep{deshpande2025trail,liu2026whowhenpro}. In contrast, SAGE localizes within a single trace whose functional checks passed and asks where the security-intent divergence first becomes trace-attributable.

\textbf{LLM-as-a-judge in software engineering.} LLM judges are increasingly used to evaluate generated code and software agents~\citep{tong2024codejudge,zhuge2025agentjudge}. However, their agreement with human evaluators varies across tasks~\citep{wang2025judge}, and their verdicts can shift with prompt formulation and answer position~\citep{zhao2026bias}. These findings are directly relevant to SAGE because origin localization uses judge-assigned assessments of expressed security reasoning. Therefore, we do not rely on a single LLM judgment as ground truth. Instead, SAGE uses prefix-only, quotation-anchored scoring and evaluates reliability through repeated scoring, cross-judge comparison, task-disjoint resampling, and human annotation of artefact-derived outputs.

\textbf{Positioning of this study.} Relative to the closest work, SAGE combines three features for localizing confirmed silent failures. First, it analyses a single recorded repair whose patch passed available syntactic and functional checks but was independently confirmed to contain a security deficiency. Second, it combines the failure anchor, prefix-only assessment of expressed security reasoning, and reconstructed artefact history to identify the earliest trace-attributable divergence under a systematic localization rule. This separates the origin turn from code provenance and, where one exists, the introducing write; it also allows origins to be assigned for inherited vulnerabilities. Third, it evaluates reliability from complementary sources: repeated scoring and cross-judge comparison for origin type and origin turn, and human annotation plus task-disjoint resampling for artefact-derived outputs. In this way, the study moves beyond asking whether an agentic repair succeeded or failed, and instead examines where a confirmed security failure first becomes observable in the repair process.

\section{Methodology}
\label{sec:methodology}

\subsection{Research Design Overview}
\label{sec:overview}
\begin{figure}[t]
    \centering
    \includegraphics[width=\linewidth]{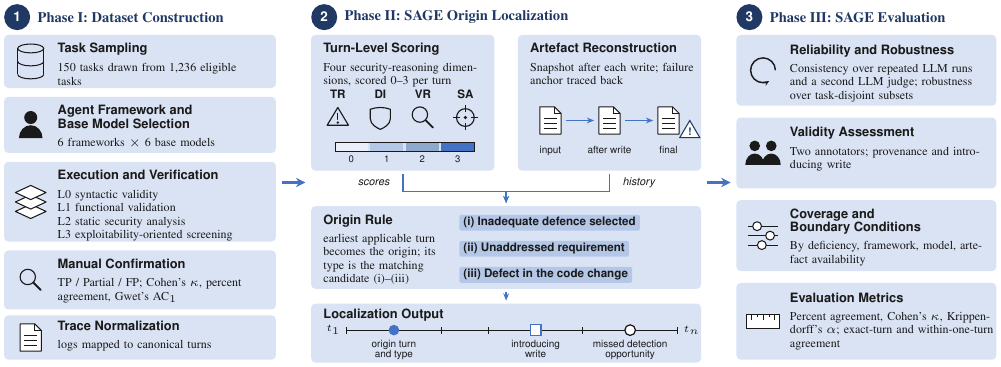}
    \caption{Overview of the research design.}
    \label{fig:overview}
\end{figure}

The study consists of three phases (see Figure~\ref{fig:overview}). Phase I constructs the corpus of confirmed silent failures, Phase II applies SAGE to localize their origins, and Phase III assesses the consistency of the localization outputs and the agreement of the artefact-derived outputs with human annotation.

\subsection{Phase I: Silent Failure Dataset Construction}
\label{sec:phase1}

\subsubsection{Task Sampling}
\label{sec:tasks}

We use two publicly available security-focused datasets: SecurityEval~\cite{siddiq2022securityeval} and CVEfixes~\cite{bhandari2021cvefixes}. Tasks were selected according to four criteria: security relevance, availability of Common Weakness Enumeration (CWE) annotations, Python implementation, and compatibility with the selected agent frameworks. After language and format filtering, the two datasets provide a pool of 1,236 eligible tasks: 121 from SecurityEval and 1,115 from CVEfixes. From this pool, we draw 150 tasks (75 per dataset), using a fixed random seed and stratification by CWE family with proportional allocation within each dataset. Each SecurityEval task supplies the code context of its prompt as the input file, and each CVEfixes task supplies the vulnerable version of the file together with a repair instruction naming the target CVE and CWE.

\subsubsection{Agent Framework and Base Model Selection}
\label{sec:systems}

We select six open-source frameworks that had active release lines at the time of corpus construction and were applicable to vulnerability repair, directly or through adaptation (see Table~\ref{tab:frameworks}). Each framework is executed with six base models: GPT-5.5~\cite{openai2026gpt55}, Claude Opus 4.8~\cite{anthropic2026opus48}, Gemini 3.1 Pro~\cite{googledeepmind2026gemini31pro}, DeepSeek-V4-Pro~\cite{deepseekai2026v4}, Kimi K3~\cite{moonshotai2026kimik3}, and Qwen3.8-Max~\cite{qwen2026qwen38max}. These models were selected for coding and reasoning capability, programmatic access, and compatibility with the frameworks' context-length and tool-use requirements.

\begin{table}[t]
\centering
\caption{Selected agent frameworks.}
\label{tab:frameworks}
\footnotesize
\begingroup
\setlength{\tabcolsep}{3pt}
\renewcommand{\arraystretch}{1.05}
\begin{tabular}{@{}
    p{0.20\columnwidth}
    p{0.07\columnwidth}
    p{0.22\columnwidth}
    p{0.47\columnwidth}
@{}}
\hline
Framework & Type & Architecture & Selection rationale \\
\hline
CAMEL~\cite{li2023camel}
& Multi & Two-role dialogue
& Planner--engineer roles with explicit handoffs \\
AG2~\cite{wu2024autogen}
& Multi & Group chat
& Round-robin chat with a security reviewer \\
Deep Agents~\cite{langchain2026deepagents}
& Multi & Planner--subagent
& Hierarchical decomposition and delegation \\
Aider~\cite{gauthier2024aider}
& Single & Edit--feedback loop
& Diff edits with lint and test feedback \\
MiniSWE-Agent~\cite{yang2024swe}
& Single & ReAct loop
& Reasoning and tool use, no inter-agent messages \\
OpenHands~\cite{wang2024openhands}
& Single & Edit--feedback loop
& Sandboxed file editing and shell execution \\
\hline
\end{tabular}
\endgroup
\end{table}

\subsubsection{Experimental Setup}
\label{sec:setup}

To limit variation arising from execution settings, all frameworks and models use the same generation parameters where supported (temperature 1.0, 600-second timeout), and each run executes in an isolated environment allocated 5 CPU cores, 16~GB of RAM, and 512~GB of storage. Task content is kept unchanged across frameworks; only its format is adapted to each framework's input interface.

\subsubsection{Execution and Verification}
\label{sec:execution}

Each task is executed once with each of the 36 framework--model configurations (5,400 runs), and each execution produces a generated patch, a complete agent trace, and any intermediate or auxiliary artefacts created during the repair process. Runs that terminate without producing a valid modification, including crashes, timeouts, and incomplete generations, are recorded as invalid traces and excluded. After this exclusion, 3,684 traces (68.22\%) enter verification, of which 331 are flagged as silent failure candidates; Table~\ref{tab:corpus} reports these counts and the excluded runs by framework. 

Generated patches are evaluated using a multi-level verification framework \cite{bai2026passingtestshidesvulnerabilities}:

\begin{itemize}
    \item \textbf{Level 0 (L0): Syntactic validity.} The generated code can be parsed without syntax errors.
    \item \textbf{Level 1 (L1): Functional validation.} The generated artefact passes the available functional checks: its imports resolve, the functions and classes defined in the input file are preserved, and the associated tests pass where the task provides them. As SecurityEval provides no test suites, its tasks are assessed with the first two checks.
    \item \textbf{Level 2 (L2): Static security analysis.} Bandit analyses the generated artefacts, with findings mapped to CWE identifiers. Patches containing high-severity findings are flagged.
    \item \textbf{Level 3 (L3): Exploitability-oriented screening.} A rule-based score $R \in [0, 100]$ is computed for each patch by pattern matching. It sums weights for attacker-controllable input sources (2--4) and dangerous sinks such as dynamic code execution or unsafe deserialization (6--12); a sink weight is multiplied by 1.5 when a source occurs in the same patch and by 1.3 when the sink matches the CWE targeted by the task. The score increases by 10 if the targeted weakness pattern persists and decreases by 2 for each detected safeguard category (at most 10). A patch is flagged when $R \geq 15$, using the maximum file score for multi-file patches. These settings follow prior work~\cite{bai2026passingtestshidesvulnerabilities}; because every flagged candidate is reviewed manually, the threshold sets screening sensitivity.
\end{itemize}
\begin{table}[t]
\centering
\caption{Silent Failure Dataset construction by framework.}
\label{tab:corpus}
\small
\begin{tabular}{lrrrrr}
\hline
Framework & Excluded runs & Valid traces & Passed L0--L3 & Failed L0/L1 & SF candidates  \\
\hline
AG2 & 337 & 563 & 459 & 49 & 55  \\
Aider & 156 & 744 & 624 & 58 & 62  \\
CAMEL & 232 & 668 & 345 & 292 & 31  \\
Deep Agents & 378 & 522 & 443 & 29 & 50  \\
MiniSWE-Agent & 383 & 517 & 424 & 27 & 66  \\
OpenHands & 230 & 670 & 549 & 54 & 67  \\
\hline
Total & 1,716 & 3,684 & 2,844 & 509 & 331  \\
\hline
\end{tabular}
\end{table}
Patches that pass L0 and L1 but are flagged at L2 or L3 are treated as silent failure candidates. CAMEL accounts for 292 of the 509 L0/L1 failures, of which 241 failed functional validation (L1), mostly because the delivered file omitted functions or classes defined in the input file. These levels provide screening signals. Only manually confirmed candidates are included in the corpus.

\subsubsection{Manual Confirmation}
\label{sec:confirmation}

For each silent failure candidate, the annotator examines the task description, generated artefact, security findings, and relevant intermediate artefacts. Candidates are classified as follows:

\begin{itemize}
    \item \textbf{True positive (TP)}: the patch retains or introduces a clear vulnerability relevant to the task;
    \item \textbf{Partial}: the patch contains a relevant mitigation attempt, but a plausible vulnerability remains; 
    \item \textbf{False positive (FP)}: the reported finding does not represent a relevant security weakness in context.
\end{itemize}

Both TP and Partial cases are retained in the confirmed silent failure corpus, whereas FP cases are excluded. To assess the consistency of this confirmation process and limit potential individual bias, we conduct an Inter-Rater Reliability (IRR) analysis. The first annotator (R1) reviews all silent failure candidates, while a second annotator (R2) independently reviews a random subset of 60 candidates using the same criteria. Agreement is assessed primarily using Cohen's $\kappa$~\cite{cohen1960coefficient}, with percent agreement and Gwet's AC$_1$~\cite{gwet2008computing} reported as complementary measures in Table~\ref{tab:irr}. As inclusion in the corpus is a binary decision (TP or Partial versus FP), we also report binary agreement: the annotators agreed on 54 of the 60 candidates (90.00\%; $\kappa = 0.74$). All disagreements were resolved through discussion between the two annotators. The resulting corpus contains 95 confirmed silent failure cases from 19 tasks (87 cases from 15 CVEfixes tasks and 8 from 4 SecurityEval tasks), each with a failure anchor recording the affected artefact and the confirmed deficiency (an absent defence, an insufficient defence, or a vulnerability introduced by the change). The largest task, a CVEfixes repair of a server-side request forgery vulnerability, contributes 16 cases.

\begin{table*}[t]
\centering
\caption{Inter-rater reliability on the silent failure candidates subset.}
\label{tab:irr}
\small

\begin{minipage}[t]{0.48\textwidth}
\centering
\textbf{(a) Inter-rater agreement}

\vspace{3pt}
\begin{tabular*}{\linewidth}{@{\extracolsep{\fill}}lc@{}}
\hline
Metric & Value \\
\hline
Observed agreement
    & 0.88  \\
Cohen's $\kappa$
    & 0.73  \\
Gwet's AC$_1$
    & 0.85  \\
\hline
\end{tabular*}
\end{minipage}
\hfill
\begin{minipage}[t]{0.48\textwidth}
\centering
\textbf{(b) Category confusion matrix ($n=60$)}

\vspace{3pt}
\begin{tabular*}{\linewidth}{@{\extracolsep{\fill}}lccc@{}}
\hline
\multicolumn{1}{c}{R1 $\backslash$ R2}
    & TP & Partial & FP \\
\hline
TP      & 6 & 1 & 3  \\
Partial & 0 & 6 & 1  \\
FP      & 1 & 1 & 41 \\
\hline
\end{tabular*}
\end{minipage}

\end{table*}

\subsubsection{Trace Normalization}
\label{sec:canonical}

The selected frameworks record messages, tool calls, and artefact changes in different formats and at different levels of granularity. To enable consistent analysis across frameworks, we normalize each execution trace into a sequence of canonical turns.

A canonical turn consists of an agent-authored message together with the tool calls, tool results, and artefact changes associated with that message. Depending on the framework and the execution step, a canonical turn may include: agent messages; handoff records; planning artefacts; review comments; intermediate code; and auxiliary artefacts.

\subsection{Phase II: SAGE Origin Localization}
\label{sec:phase2}

In Phase~II, SAGE operates on the canonical turns produced by trace normalization. It first measures the security reasoning expressed in every turn and then locates two positions for each confirmed silent failure: the introducing write, derived from the artefact history alone, and the origin turn, derived from the artefact history together with the per-turn scores.

\subsubsection{SAGE Turn-Level Scoring}
\label{sec:sage}

SAGE measures the security reasoning expressed by an agent over the course of an execution. For each canonical turn, SAGE independently scores the four dimensions in Table~\ref{tab:sage} on a common scale. 

Alongside these dimensions, each turn is assigned two auxiliary labels: (i) \textbf{Applicability}, which distinguishes substantive planning, implementation, review, and verification turns from administrative framework events, such as environment setup or tool polling; and (ii) \textbf{Security opportunity}, which indicates whether the turn makes or reviews a decision, code path, or artefact materially connected to the task's security intent, so that security reasoning at that turn could affect whether this intent is met. The judge assigns this label from the task context and the trace prefix only, without access to the confirmed deficiency; the origin rule links candidate turns to the deficiency afterwards (see Section~\ref{sec:localization}).

\begin{table*}[t]
\centering
\caption{The SAGE instrument: the four scoring dimensions and the common anchored scale applied to each of them.}
\label{tab:sage}
\small
\begin{minipage}[t]{0.56\textwidth}
\centering
\textbf{(a) Scoring dimensions}\\[2pt]
\begin{tabular}{@{}lp{0.60\linewidth}@{}}
\hline
Dimension & Interpretation \\
\hline
Threat Recognition (TR) & The relevant threat, attack path, or trust boundary is identified \\
Defensive Intent (DI) & An appropriate defence and its intended security role are specified \\
Verification Rigour (VR) & The implemented defence is tested or challenged \\
Scope Awareness (SA) & The code and artefacts requiring security consideration are recognized \\
\hline
\end{tabular}
\end{minipage}
\hfill
\begin{minipage}[t]{0.40\textwidth}
\centering
\textbf{(b) Dimension scale}\\[2pt]
\begin{tabular}{@{}lp{0.74\linewidth}@{}}
\hline
Score & Expressed reasoning \\
\hline
0 & None related to security \\
1 & Generic security remarks \\
2 & Specific to the task \\
3 & Task-specific, and addresses completeness or the conditions under which
the defence holds \\
\hline
\end{tabular}
\end{minipage}
\end{table*}

We use GPT-5.6 Sol~\cite{openai2026gpt56} as the primary LLM judge because it lies outside the evaluated base-model set, so that no trace is scored by the model that produced it. The judge scores one canonical turn (see Section~\ref{sec:canonical}) at a time in trace order, receiving the task context and the trace up to and including that turn; later turns, multi-level verification results, and confirmation labels are withheld. To ensure auditability, every non-zero score must include a verbatim quotation linked to a canonical item and, through it, to the raw execution trace. All turns involving code planning, implementation, review, or verification are scored. An applicable turn with no expressed security reasoning receives a score of 0, indicating the absence of such reasoning in the recorded trace. Only administrative framework events, such as formatting, file bookkeeping, and framework handshake messages, are excluded.

\subsubsection{SAGE Origin Localization}
\label{sec:localization}

After scoring, SAGE localizes each confirmed silent failure using the failure anchor recorded during manual confirmation, the reconstructed artefact history, and the per-turn scores. The anchor identifies the confirmed deficiency and the affected artefact, but contains no origin label. The origin turn is the earliest canonical turn at which the available evidence indicates a failure to satisfy the task's security intent and links that failure to the confirmed deficiency under the rule below. This operational definition identifies a trace-attributable divergence. Supporting evidence may come from expressed reasoning or artefact changes; an introducing write is therefore not required to localize the origin of an inherited vulnerability.

Localization proceeds in two steps. The first step establishes code provenance. The original vulnerable artefact is treated as the baseline and compared with the snapshot produced after each write. A deficiency is classified as agent-introduced when its anchor-linked fragment is absent from the baseline and appears in a later snapshot. This category includes newly introduced weaknesses, insufficient mitigations, and the removal or weakening of an existing defence. A deficiency is classified as inherited from the input file when it already exists in the original code and is not eliminated by subsequent writes.

The second step assigns two trace positions to each case. The first is derived from the artefact history alone. For an agent-introduced vulnerability, the introducing write is the earliest write after which the vulnerability appears and remains present through the final snapshot. If an intermediate write removes the vulnerability and a later write reinstates it, the reinstating write is selected. An inherited vulnerability has no introducing write. The second position, the origin turn, is derived from the artefact history together with the per-turn SAGE scores. A fixed decision rule considers the following candidates and selects the earliest applicable turn:

\begin{enumerate}
\item[(i)] \textbf{Inadequate defence selected.} The earliest turn before the introducing write (or before the delivered repair for an inherited vulnerability) at which the agent expresses a defence linked to the failure anchor, indicated by a non-zero defensive-intent score. The score records only that a defence is expressed; the defence is treated as inadequate because manual confirmation established that the delivered patch still contains the deficiency it addresses.

\item[(ii)] \textbf{Unaddressed requirement.} The first turn, possibly the introducing write itself, at which the agent has an opportunity to address the task's security intent but expresses neither threat recognition nor defensive intent.

\item[(iii)] \textbf{Defect in the code change.} The introducing write, selected only when no turn qualifies under (i) or (ii); in that case, the agent expresses threat recognition or defensive intent at the write itself. This candidate does not apply to inherited vulnerabilities.
\end{enumerate}

The origin type therefore records which candidate of the rule was selected, while the origin turn identifies its position in the trace and may precede the introducing write. An auxiliary output flags implementation or review turns after the origin whose verification-rigour or scope-awareness score is at most 1 as missed detection opportunities; these turns are reported separately and are not selected as origins. The threshold follows the scale in Table~\ref{tab:sage}(b): a score of 0 or 1 means that the turn records no security-related verification or scope reasoning, or only generic remarks, whereas a score of 2 requires reasoning specific to the task.

After applying the rule, each case is assigned one of two outcomes: (i) Localized indicates that an origin turn was selected. (ii) Undetermined indicates that no origin turn was selected, for one of two recorded reasons: no candidate turn could be linked to the confirmed deficiency, or the per-turn scores required by the rule are unavailable because scoring failed for the trace. Provenance and the introducing write do not depend on the scores and are reported for both Localized and Undetermined outcomes.

All localization results are checked automatically. For an agent-introduced deficiency, the check verifies that the anchor-linked fragment is absent from the baseline, is introduced or reintroduced by the selected write, and remains present thereafter. For a deficiency inherited from the input file, the check verifies that the anchored deficiency is present in the baseline and persists across subsequent snapshots. These checks ensure the fragment history aligns with the assigned provenance and that the origin turn precedes or coincides with the introducing write. Fragment matching only locates code; the vulnerability itself is established through manual confirmation (see Section \ref{sec:confirmation}).

\subsection{Phase III: SAGE Evaluation}
\label{sec:evaluation}
\begin{figure}[t]
    \centering
    \includegraphics[width=\linewidth]{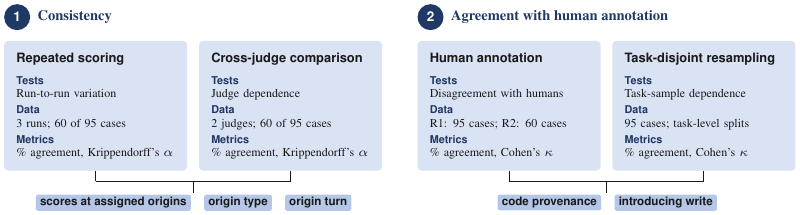}
    \caption{Multi-source evaluation design of SAGE.}
    \label{fig:eva}
\end{figure}
In Phase III, we assess SAGE through a multi-source reliability evaluation (see Figure~\ref{fig:eva}): repeated scoring and cross-judge comparison assess the consistency of judge-dependent outputs, while human annotation and task-disjoint resampling assess agreement on artefact-derived outputs and its sensitivity to task composition.

\subsubsection{Reliability and Robustness}
\label{sec:reliability}

We evaluate the reliability and robustness of SAGE through three complementary analyses:

\begin{enumerate}
\item[(i)] \textbf{Scoring stability.} The primary LLM judge (GPT-5.6 Sol) scores the full set of confirmed silent failures once. A stratified random subset (60 of the 95 SF cases, across all frameworks) is scored in two additional runs, giving three scores per case in the subset. The subset covers different frameworks and reconstruction conditions. Across the three runs, we evaluate consistency on the four dimension scores at the origin turn and on the origin type and origin turn re-derived from each run's scores.

\item[(ii)] \textbf{Cross-judge consistency.} A second LLM judge (Claude Opus 5 \cite{anthropic2026opus5}) independently scores the same subset using the same rubric, prompt, and prefix-only context. This judge is blind to the primary scores, human annotations, and localization results. We compare the two judges on the same measures.

\item[(iii)] \textbf{Robustness across data subsets.} To assess whether SAGE--human agreement depends on the composition of the evaluation set, we conduct two task-level resampling analyses on the 95 confirmed SF cases. First, across 1,000 random partitions, tasks are divided into two approximately equal, task-disjoint subsets, keeping all cases from the same task within the same subset. Agreement is recomputed separately for each resulting subset using the same metrics. Second, a leave-one-task-out analysis recomputes agreement after iteratively excluding each task.
\end{enumerate}

\subsubsection{Validity Assessment}
\label{sec:validity}

We compare SAGE's code provenance and introducing-write labels with human annotation; the origin type and origin turn are not annotated by humans and are assessed for consistency only (see Section~\ref{sec:reliability}). The first annotator (R1) labels all 95 confirmed cases, and a second annotator (R2) independently labels a stratified subset of 60 cases. Neither annotator is shown the SAGE scores or automated localization outputs. Both use the same failure anchors and reconstructed artefact histories as SAGE to label code provenance and the introducing write (a specific write turn, no introducing write, or cannot determine). Because R1 participated in corpus construction and in the design of the localization rule, R1 may be familiar with the cases and with the rule's logic; R2 was not involved in either step.

\subsubsection{Localization Coverage and Boundary Conditions}
\label{sec:coverage-boundaries}

We examine localization coverage and its boundary conditions through two analyses:

\begin{enumerate}
\item[(i)] \textbf{Trace evidence and localizability.} For each valid trace, we record the available evidence sources, number of canonical turns, number of applied writes, and reconstruction condition. For each confirmed silent failure, we record its localization outcome, code provenance, and supporting evidence.

\item[(ii)] \textbf{Boundary conditions.} We examine how SAGE localization outcomes vary across failure categories, agent frameworks, base models, reconstruction conditions, and artefact availability. For each condition, we report localizability and the proportion of Undetermined cases. Where data permit, we also compare scoring consistency and cross-judge agreement. Results based on small groups are interpreted with caution.
\end{enumerate}

\subsubsection{Evaluation Metrics}
\label{sec:statistics}

To evaluate the reliability and validity of SAGE localization, we report joint localization coverage and, among jointly localized cases, agreement on origin type and exact or within-one-turn origin localization. Krippendorff's $\alpha$ \cite{hayes2007answering} assesses agreement on origin type and turn (cases with at least two ratings) and dimension scores at assigned origins (ordinal). Human comparisons use percent agreement and Cohen's $\kappa$ for code provenance and introducing-write labels, comparing annotators with each other and SAGE with each annotator separately.
\section{Results}
\label{sec:results}
\begin{table*}[t]
\centering
\caption{Overview of SAGE localization results ($N=95$ confirmed silent failures).}
\label{tab:rq1-overview}
\small
\renewcommand{\arraystretch}{1.10}

\begin{minipage}[t]{0.48\textwidth}
\vspace{0pt}
\centering
\textbf{(a) Code provenance in silent failures}

\vspace{3pt}
\begin{tabular*}{\linewidth}
{@{\extracolsep{\fill}}lrr@{}}
\hline
Outcome & $N$ & \% \\
\hline
Inherited from the input file
    & 73 & 76.84 \\
Introduced by the agent
    & 19 & 20.00 \\
No identifiable code location
    & 3 & 3.16 \\
\hline
\end{tabular*}
\end{minipage}
\hfill
\begin{minipage}[t]{0.48\textwidth}
\vspace{0pt}
\centering
\textbf{(b) Origin type assigned by SAGE}

\vspace{3pt}
\begin{tabular*}{\linewidth}
{@{\extracolsep{\fill}}lrr@{}}
\hline
Outcome & $N$ & \% \\
\hline
Unaddressed requirement
    & 51 & 53.68 \\
Inadequate defence selected
    & 37 & 38.95 \\
Defect in the code change
    & 5 & 5.26 \\
Undetermined
    & 2 & 2.11 \\
\hline
\end{tabular*}
\end{minipage}

\vspace{0.8em}

\begin{minipage}[t]{0.48\textwidth}
\vspace{0pt}
\centering
\textbf{(c) Origin timing for agent-introduced code
($n=19$)}

\vspace{3pt}
\begin{tabular*}{\linewidth}
{@{\extracolsep{\fill}}lrr@{}}
\hline
Outcome & $n$ & \% \\
\hline
Before code introduction
    & 14 & 73.68 \\
At code introduction
    & 5 & 26.32 \\
\hline
\end{tabular*}
\end{minipage}
\hfill
\begin{minipage}[t]{0.48\textwidth}
\vspace{0pt}
\centering
\textbf{(d) Missed opportunities for later detection}

\vspace{3pt}
\begin{tabular*}{\linewidth}
{@{\extracolsep{\fill}}lrr@{}}
\hline
Outcome & $N$ & \% \\
\hline
At least one missed opportunity recorded
    & 59 & 62.11 \\
No missed opportunity recorded
    & 36 & 37.89 \\
\hline
\end{tabular*}
\end{minipage}
\end{table*}

\subsection{Trace Evidence and Localized Origins (RQ1)}
\label{sec:results-rq1}

To answer RQ1, we examine how artefact histories and reasoning records support SAGE's origin assignments, and which subsequent turns are flagged as missed detection opportunities. Table~\ref{tab:rq1-overview} summarizes code provenance, assigned origin types, origin timing relative to the introducing write, and missed detection opportunities. SAGE assigns an origin to 93 of the 95 confirmed cases (97.89\%), leaving two cases Undetermined (2.11\%).

\begin{figure}[t]
  \centering
  \includegraphics[width=\textwidth]{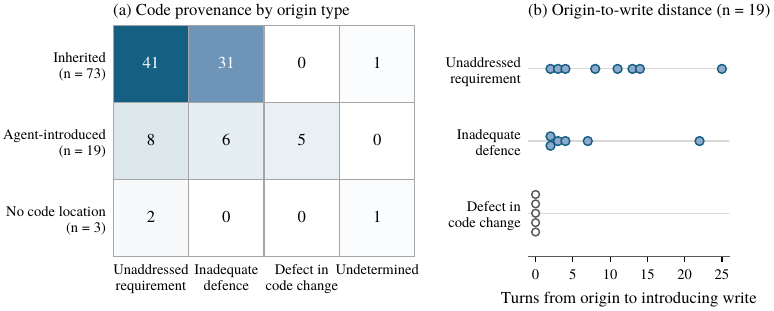}
  \caption{Origin localization outcomes. (a) Code provenance against origin type for all 95 confirmed silent failures; (b) Distance in canonical turns between the origin turn and the introducing write for the 19 cases in which the agent introduced the vulnerable code, one dot per case.}
  \label{fig:rq1-main}
\end{figure}

\textbf{Artefact histories establish code provenance.} SAGE classified 73 cases as inherited from the input file (76.84\%) and 19 as agent-introduced (20.00\%); three cases could not be linked to an identifiable code location (3.16\%; see Table~\ref{tab:rq1-overview}(a)). These outputs describe the available code-history evidence, while localization of the origin additionally draws on the trace record. For example, in one case, the agent completed the requested repair but introduced an unsafe configuration absent from the input file. The artefact change provides evidence for an introducing write, while identifying any earlier origin requires examining the preceding reasoning. In another case, a vulnerable database operation remained unchanged from the input file to the final patch. Its history supports inherited provenance but provides no introducing write, so origin candidates must be sought in the reasoning record.

Beyond code changes, the agents' repair plans and explanations reveal how vulnerabilities persisted or emerged. SAGE attributed the origin to an unaddressed requirement in 51 cases (53.68\%), the selection of an inadequate defence in 37 (38.95\%), and a defect in the code change in 5 (5.26\%) (see Table~\ref{tab:rq1-overview}(b)). Inherited vulnerabilities accounted for most unaddressed-requirement origins (41/51, 80.39\%) and inadequate-defence origins (31/37, 83.78\%; see Figure~\ref{fig:rq1-main}(a)).

\textbf{In most agent-introduced cases, the assigned origin preceded the introducing write.} Among the 19 cases in which the agent introduced the vulnerability (see Table~\ref{tab:rq1-overview}(c)), the identified origin turn preceded the introducing write in 14 cases (73.68\%). Of these, SAGE assigned eight origins to an unaddressed security requirement and six to the selection of an inadequate defence (see Figure~\ref{fig:rq1-main}(a)). In the remaining five cases (26.32\%), the origin coincided with the introducing write; all five were single-message Aider traces in which the patch was generated and written in one step, leaving no earlier recorded turn for localization. Because the rule restricts origin candidates to the introducing write or earlier turns, this temporal ordering holds by construction. SAGE also assigned an origin turn in nearly all cases without an identifiable introducing write (74/76, 97.37\%). These cases comprise vulnerabilities inherited from the input file and cases for which no specific code location could be identified. 

\textbf{SAGE flagged missed detection opportunities in 59 cases.} SAGE records a post-origin implementation or review turn as a missed detection opportunity when its verification-rigour or scope-awareness score is at most 1 (see Section~\ref{sec:localization}). At least one such opportunity was identified in the majority of cases (59/95, 62.11\%) (see Table~\ref{tab:rq1-overview}(d)). The flags were concentrated in frameworks with multi-turn traces (AG2 29/32, Deep Agents 13/13, MiniSWE-Agent 11/15, OpenHands 4/9, CAMEL 2/4), whereas none of the 22 Aider cases was flagged because their single-message traces contain no turn after the origin. The overall share therefore largely reflects the framework composition of the corpus. Even in AG2, whose group chat includes a dedicated security reviewer, 28 of the 32 cases had a flagged reviewer turn. These flags reflect low recorded verification rigour or scope awareness under the SAGE rule. For example, in one case, the reviewer agent endorsed the completed sorting repair and reported no remaining issues, although an unsafe code-access configuration remained. SAGE left two of the 95 cases undetermined (2.11\%; see Figure~\ref{fig:rq1-main}(a)): one with a vulnerability inherited from the input file and one without an identifiable code location. In both, no candidate turn could be linked to the confirmed deficiency. For the other two cases without an identifiable code location, SAGE assigned an origin from the planning and dialogue records alone, in both cases as an unaddressed security requirement.

\begin{keyfindingbox}{Key Findings from RQ1}
We observed that agent traces provide complementary evidence about both failure origins and potential intervention points. Artefact histories support code provenance and introducing-write identification, while reasoning records support origin assignments to unaddressed security requirements or inadequate defence choices. Later implementation and review turns with low verification rigour or scope awareness were flagged as missed detection opportunities. Together, these findings suggest that trace-attributable divergence can become observable before later code edits, while subsequent turns provide additional opportunities for intervention.
\end{keyfindingbox}

\begin{table*}[t]
\centering
\caption{Reliability of SAGE localization outputs.}
\label{tab:rq2-summary}
\small
\renewcommand{\arraystretch}{1.05}
\setlength{\tabcolsep}{3pt}

\noindent
\begin{minipage}[t]{0.485\textwidth}
\vspace{0pt}
\centering
\textbf{(a) Origin consistency}

\vspace{2pt}
\begin{tabular*}{\linewidth}
{@{\extracolsep{\fill}}lccc@{}}
\toprule
& \shortstack{Repeated\\scoring}
& \shortstack{Cross\\judge}
& \shortstack{All four\\ratings} \\
\midrule
Eligible cases ($n$)
    & 57 & 50 & 48 \\
Type agreement (\%)
    & 96.49 & 92.00 & 89.58 \\
Exact-turn agreement (\%)
    & 87.72 & 78.00 & 75.00 \\
Within $\pm 1$ turn (\%)
    & 91.23 & 84.00 & 81.25 \\
Origin type $\alpha$
    & 0.96 & 0.86 & 0.92 \\
Origin turn $\alpha$
    & 0.82 & 0.59 & 0.74 \\
\bottomrule
\end{tabular*}
\end{minipage}%
\hfill%
\begin{minipage}[t]{0.485\textwidth}
\vspace{0pt}
\centering
\textbf{(b) Agreement across task subsets}

\vspace{2pt}
\begin{tabular*}{\linewidth}
{@{\extracolsep{\fill}}lcc@{}}
\toprule
& \shortstack{Provenance\\(\%)}
& \shortstack{Introducing\\write (\%)} \\
\midrule
All 95 cases
    & 84.21 & 81.05 \\
Subset median
    & 84.21 & 81.06 \\
Subset central 95\%
    & 74.42--96.88 & 69.39--93.33 \\
Leave-one-task-out
    & 82.56--89.87 & 79.07--87.34 \\
\bottomrule
\end{tabular*}
\end{minipage}
\vspace{0.45em}

\begin{minipage}{\textwidth}
\centering
\textbf{(c) Agreement with human annotations}

\vspace{2pt}
\begin{tabular*}{\linewidth}
{@{\extracolsep{\fill}}lccccc@{}}
\toprule
Comparison & $n$
    & \multicolumn{2}{c}{Code provenance}
    & \multicolumn{2}{c}{Introducing-write label} \\
\cmidrule(lr){3-4}
\cmidrule(l){5-6}
& & Agreement (\%) & $\kappa$
  & Agreement (\%) & $\kappa$ \\
\midrule
R1 vs.\ R2
    & 60 & 90.00 & 0.78 & 75.00 & 0.58 \\
SAGE vs.\ R1
    & 60 & 88.33 & 0.74 & 85.00 & 0.70 \\
SAGE vs.\ R2
    & 60 & 85.00 & 0.66 & 78.33 & 0.62 \\
\cmidrule(l){1-6}
SAGE vs.\ R1
    & 95 & 84.21 & 0.62 & 81.05 & 0.58 \\
\bottomrule
\end{tabular*}
\end{minipage}

\vspace{2pt}
\begin{minipage}{\textwidth}
\footnotesize
\textit{Notes.}
(a) Agreement uses cases localized in all compared ratings; $\alpha$ uses cases with at least two ratings. All four ratings combine three primary-judge runs and one independent judge. (b) Subsets come from 1,000 random task-disjoint partitions; all values are agreement with R1. (c) Introducing-write labels denote a write turn, no introducing write, or cannot determine.
\end{minipage}
\end{table*}

\subsection{Reliability of SAGE Localization (RQ2)}
\label{sec:results-rq2}

\begin{figure}[t]
  \centering
  \includegraphics[width=\textwidth]{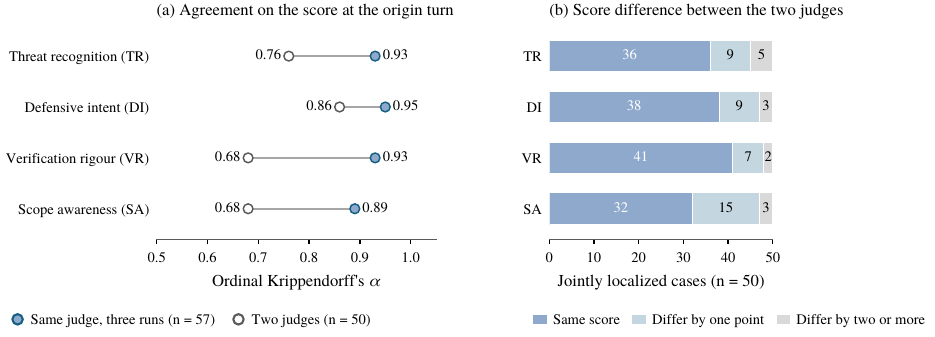}
  \caption{Reliability of the judge-assigned profile scores at the origin turn.}
  \label{fig:rq2-profile-agreement}
\end{figure}

To answer RQ2, we assess the consistency of SAGE's origin assignments across repeated scoring and independent judges, as well as the alignment of its artefact-derived outputs with human annotations across different task subsets. The repeated-scoring, cross-judge, and inter-annotator comparisons use the same stratified random subset of 60 cases. Table~\ref{tab:rq2-summary} summarizes these results.

Repeated scoring suggests that SAGE reproduces the type of the recorded divergence more consistently than its exact turn. Table~\ref{tab:rq2-summary}(a) shows that the primary judge assigned an origin in all three runs for most cases (57/60, 95.00\%). Within this shared set, agreement was higher for origin type (55/57, 96.49\%) than for the exact origin turn (50/57, 87.72\%), suggesting that the nature of the recorded divergence was more consistently reproduced than its precise position. Allowing a one-turn difference increased agreement to 91.23\% (52/57), covering two additional cases whose assigned origins fell on neighbouring turns. Krippendorff's $\alpha$ followed the same pattern, reaching 0.96 for origin type and 0.82 for origin turn. Figure~\ref{fig:rq2-profile-agreement}(a) complements these results by showing consistent scoring across all four SAGE dimensions, with ordinal $\alpha$ ranging from 0.89 to 0.95. Ordinal $\alpha$ between the two judges was nevertheless lower than within the primary judge: 0.76 for threat recognition, 0.86 for defensive intent, and 0.68 for both verification rigour and scope awareness. The latter two dimensions also determine which turns are flagged as missed detection opportunities, so these flags may likewise be sensitive to the choice of judge.

Cross-judge comparison showed the same contrast between origin type and exact turn, with a wider gap than under repeated scoring. Among cases localized by both judges (50/60, 83.33\%), origin-type agreement (46/50, 92.00\%) exceeded exact-turn agreement (39/50, 78.00\%; Table~\ref{tab:rq2-summary}(a)). Allowing a one-turn difference increased agreement to 84.00\% (42/50), indicating that neighbouring turns accounted for some, but not all, of the temporal disagreements. The same pattern held across all four ratings, comprising the primary judge's three runs and the independent judge: $\alpha$ was higher for origin type (0.92) than for origin turn (0.74). At the level of individual dimensions, Figure~\ref{fig:rq2-profile-agreement}(b) shows that scores differed by at most one point in 90.00\%--96.00\% of jointly localized cases, suggesting broad agreement in the underlying assessments despite variation in exact turn selection. 

SAGE showed stronger agreement with the annotators on code provenance than on introducing-write labels. On the shared 60-case subset, agreement with each annotator ranged from 85.00\% to 88.33\% for provenance and from 78.33\% to 85.00\% for the introducing-write label. Table~\ref{tab:rq2-summary}(c) places this alignment alongside the agreement between annotators, which was higher for provenance (54/60, 90.00\%; $\kappa = 0.78$) than for the introducing-write label (45/60, 75.00\%; $\kappa = 0.58$). This suggests that determining whether a vulnerability was inherited or introduced was more reproducible than identifying its introducing write. Where the annotators reached consensus, SAGE matched their labels in most cases, both for provenance (49/54, 90.74\%) and for the introducing-write label (44/45, 97.78\%). 

However, although overall alignment with the human reference remains good, agreement rates varied across task subsets, indicating that aggregate estimates may depend on sample composition. As shown in Table~\ref{tab:rq2-summary}(b), across 1,000 random task-disjoint partitions, we computed subset-level agreement for code provenance (median 84.21\%; central 95\% of values: 74.42\%--96.88\%) and for the introducing write (median 81.06\%; central 95\% of values: 69.39\%--93.33\%). In the leave-one-task-out analysis, we recomputed agreement after excluding each task in turn, for code provenance (82.56\%--89.87\%) and for the introducing write (79.07\%--87.34\%). The highest values for both labels followed the exclusion of the task with the most cases (16 of 95), which accounted for a large share of the disagreements on code provenance (7 of 15) and on the introducing write (8 of 18). Overall, subset-level estimates appear to vary considerably with task composition, whereas no single task appears to inflate the aggregate estimate.

\begin{keyfindingbox}{Key Findings from RQ2}
We found that SAGE identified the origin type more consistently than the exact origin turn, with lower agreement between judges than across repeated scoring. For the artefact-derived outputs, agreement with human annotators was higher for code provenance than for introducing-write labels and varied across task subsets. Together, these results capture different aspects of SAGE's localization and are best interpreted jointly.
\end{keyfindingbox}

\subsection{Variation and Boundary Conditions (RQ3)}
\label{sec:results-rq3}
\begin{figure*}[t]
\centering
\includegraphics[width=\textwidth]{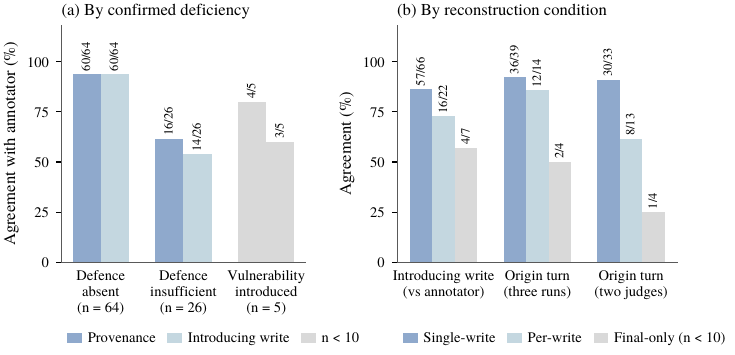}
\caption{Agreement by confirmed deficiency and by reconstruction condition.}
\label{fig:rq3-main}
\end{figure*}

To answer RQ3, we analyse how localization reliability varies with the kind of confirmed deficiency, the agent framework, the base model, and the available artefact history. We compare SAGE with R1 on code provenance and the introducing-write label, and use repeated scoring and independent judges to assess the stability of the reasoning-based origin turn. Figure~\ref{fig:rq3-main} and Table~\ref{tab:rq3-groups} summarize the observed differences. All subgroup comparisons are descriptive because the confirmed SF subsets differ in task, vulnerability, and trace composition. Estimates are particularly limited for CAMEL ($n=4$), OpenHands ($n=9$), Kimi K3 ($n=5$), and Qwen3.8-Max ($n=1$).

SAGE aligned most with human references in absent-defence cases, but this advantage did not improve origin-turn stability. Figure~\ref{fig:rq3-main}(a) groups the 95 cases according to manual confirmation of the delivered code: an absent defence, an insufficient defence, or a vulnerability introduced by the change. This classification describes the confirmed silent failure and differs from the origin types reported in Section~\ref{sec:results-rq1}. For absent-defence cases, SAGE agreed with R1 on both provenance and the introducing-write label in 93.75\% of cases (60/64). Agreement was notably lower for insufficient-defence cases, reaching 61.54\% (16/26) for provenance and 53.85\% (14/26) for the introducing-write label. The high agreement for absent defences largely reflects their inherited provenance, for which no introducing write exists within the observed trace. However, the repeated scoring reveals a different pattern: all consistently localized insufficient-defence cases retained the same origin turn (16/16, 100.00\%), while the assigned turn changed in six absent-defence cases. This contrast suggests that agreement with R1 on the artefact-derived outputs and origin-turn consistency under repeated scoring capture different properties of the localization.

\begin{table*}[t]
\centering
\caption{Agreement between the localization output and annotator on the 95 confirmed SF cases, by agent framework (a) and by base model (b). $^{*}$\,Groups with fewer than ten cases.}
\label{tab:rq3-groups}
\small
\setlength{\tabcolsep}{3pt}
\noindent
\begin{minipage}[t]{0.48\textwidth}
\vspace{0pt}
\centering
\textbf{(a) Agent frameworks}
\vspace{2pt}

\begin{tabular*}{\linewidth}{@{\extracolsep{\fill}}lrrr@{}}
\hline
Framework & $n$ & Provenance & Introducing write \\
\hline
CAMEL         &  4\rlap{$^{*}$} &  4 (100.00\%) &  4 (100.00\%) \\
AG2           & 32 & 28 (87.50\%)  & 28 (87.50\%)  \\
Deep Agents   & 13 &  9 (69.23\%)  &  9 (69.23\%)  \\
Aider         & 22 & 18 (81.82\%)  & 18 (81.82\%)  \\
MiniSWE-Agent & 15 & 13 (86.67\%)  & 11 (73.33\%)  \\
OpenHands     &  9\rlap{$^{*}$} &  8 (88.89\%)  &  7 (77.78\%)  \\
\hline
\end{tabular*}
\end{minipage}%
\hfill%
\begin{minipage}[t]{0.48\textwidth}
\vspace{0pt}
\centering
\textbf{(b) Base models}
\vspace{2pt}

\begin{tabular*}{\linewidth}{@{\extracolsep{\fill}}lrrr@{}}
\hline
Base model & $n$ & Provenance & Introducing write \\
\hline
Claude Opus 4.8 & 21 & 16 (76.19\%)  & 16 (76.19\%)  \\
GPT-5.5         & 16 & 10 (62.50\%)  & 10 (62.50\%)  \\
DeepSeek-V4-Pro & 24 & 22 (91.67\%)  & 21 (87.50\%)  \\
Gemini 3.1 Pro  & 28 & 27 (96.43\%)  & 25 (89.29\%)  \\
Kimi K3         &  5\rlap{$^{*}$} &  5 (100.00\%) &  5 (100.00\%) \\
Qwen3.8-Max    &  1\rlap{$^{*}$} &  0 (0.00\%)   &  0 (0.00\%)   \\
\hline
\end{tabular*}
\end{minipage}
\end{table*}

For agent frameworks, differences emerge both in human alignment and in the opportunities available to detect vulnerabilities down the trace. Table~\ref{tab:rq3-groups}(a) shows that AG2 and Aider maintained identical within-framework agreement for provenance and introducing-write labels (87.50\%, 28/32; and 81.82\%, 18/22, respectively). However, for MiniSWE-Agent, provenance agreement (13/15, 86.67\%) exceeded introducing-write agreement (11/15, 73.33\%), indicating that consensus on code origin does not necessarily extend to the specific write operation. Interaction architectures were also associated with differences in the available subsequent evidence: AG2 traces frequently featured reviewer turns marked as missed detection opportunities (28/32, 87.50\%), whereas the 22 single-message Aider cases provided no subsequent turns for evaluation. These differences suggest that an agent framework's trace structure affects what evidence is available for localization.

High repeatability in SAGE's origin assignments did not always coincide with the human reference across base models. Among the larger groups in Table~\ref{tab:rq3-groups}(b), SAGE achieved the highest human agreement on traces generated by Gemini 3.1 Pro for both code provenance (27/28, 96.43\%) and introducing-write labels (25/28, 89.29\%), followed by DeepSeek-V4-Pro. Conversely, SAGE showed lower human agreement on GPT-5.5 agent traces (10/16, 62.50\% for both labels); yet across all three repeated runs, SAGE assigned the same origin turn to every consistently localized GPT-5.5 case (10/10, 100.00\%). This contrast illustrates that high internal consistency can coexist with disagreement over the underlying code evidence. 

\textbf{Localization agreement was lowest when intermediate artefact states were unavailable.} As shown in Figure~\ref{fig:rq3-main}, final-only traces, which provide no intermediate code states, had the lowest agreement across all three evaluations. In the human comparison, introducing-write agreement was 57.14\% (4/7) for final-only traces, compared with 72.73\% (16/22) for per-write traces and 86.36\% (57/66) for single-write traces. The same pattern appeared for the reasoning-based origin turn. Among jointly localized cases, exact-turn agreement for final-only traces was 50.00\% (2/4) across repeated scoring and 25.00\% (1/4) across judges, compared with 92.31\% (36/39) and 90.91\% (30/33), respectively, for single-write traces. Although the final-only groups were small, their consistently lower agreement suggests that the absence of intermediate artefact states limits the evidence available for assigning a precise origin turn, which points to the value of preserving such states.

\begin{keyfindingbox}{Key Findings from RQ3}
We observed that localization reliability varied with trace content, artefact granularity, and the nature of the confirmed silent failure. Final-only traces showed the lowest temporal agreement across all three evaluations, suggesting that intermediate artefact states provide useful evidence for assigning precise origin turns. Repeatability and human alignment also captured complementary aspects of reliability: stable origin assignments could coexist with lower agreement with the human reference. Finally, absent-defence cases showed higher agreement on artefact-derived outputs, while insufficient defences were associated with greater disagreement over code provenance and introducing-write labels. Differences across agent frameworks and base models were also observed, although these estimates reflect the composition of the confirmed-failure subsets within each group.
\end{keyfindingbox}

\section{Discussion}
\label{sec:discussion}

\subsection{Implications}
\label{sec:implications}

\textbf{Final-code analysis and origin localization answer different questions (RQ1).} The final patch shows that a vulnerability is present, and the artefact history adds its code provenance and, where one exists, the introducing write. Neither shows where the repair process first diverged from the task's security intent. In 73 of the 95 cases (76.84\%), the vulnerable code was inherited from the input file (see Table~\ref{tab:rq1-overview}(a)): a diff reports no change at the vulnerable location, so code-level analysis has no position to report. SAGE nevertheless assigned an origin in 72 of these cases from the recorded reasoning (see Figure~\ref{fig:rq1-main}(a)). In the 19 agent-introduced cases, a diff does identify the introducing write, but in 14 of them (73.68\%) the origin was an earlier turn, as an unaddressed requirement (8) or an inadequate defence choice (6) (see Table~\ref{tab:rq1-overview}(c) and Figure~\ref{fig:rq1-main}). In these cases, the write marks where the vulnerability entered the file, whereas the divergence was already observable at an earlier turn. Code-level analysis would thus report an introducing write in 19 of the 95 cases, and this write was the assigned origin in only five, all from single-message traces with no earlier turn (see Table~\ref{tab:rq1-overview}(b)).

\textbf{Origin type and exact origin turn should be evaluated separately (RQ2).} Agreement on the origin type exceeded agreement on the exact origin turn under repeated scoring (96.49\% against 87.72\%) and between judges (92.00\% against 78.00\%) (see Table~\ref{tab:rq2-summary}(a)). The two outputs depend on the scores in different ways. The origin type records which candidate of the rule was selected, whereas the exact turn also depends on the score assigned at each individual turn, so a small score difference at one turn may move the origin while leaving its type unchanged. Figure~\ref{fig:rq2-profile-agreement}(b) is consistent with this reading: the dimension scores of the two judges differed by at most one point in 90.00\% to 96.00\% of the jointly localized cases, yet the judges selected the same turn in 78.00\%. A one-turn tolerance recovered only part of the difference, raising agreement to 91.23\% and 84.00\%, so some disagreements extend beyond neighbouring turns. A single localization score would hide this contrast. We therefore suggest reporting the origin type and the origin turn separately, and the turn both exactly and within one turn. The origin type may remain informative, for example about whether the divergence concerns requirement coverage or defence selection, when the exact turn is uncertain. These estimates should also be read alongside joint localization coverage, since agreement among jointly localized cases does not describe cases left undetermined in one or more assessments.

\textbf{Human agreement and repeated scoring measure different things (RQ2, RQ3).} The two forms of evidence concern different outputs. Human annotation covers code provenance and the introducing write, which are derived from the artefact history and do not depend on the judge scores. Repeated scoring and cross-judge comparison cover the origin type and the origin turn, which do. Agreement with the annotators therefore indicates whether the artefact-derived outputs match a human reading of the code history, and repeated scoring indicates whether the judge-dependent outputs are stable; neither shows that a reasoning-level origin is correct. Our results show that the two can diverge. In the GPT-5.5 group, SAGE assigned the same origin turn in all three runs to each of the ten cases localized in every run, while agreement with R1 was 10/16 on both labels (see Table~\ref{tab:rq3-groups}(b) and Section~\ref{sec:results-rq3}). Similarly, insufficient-defence cases had the lowest agreement with R1 (16/26 and 14/26; see Figure~\ref{fig:rq3-main}(a)) but an identical origin turn across runs in all 16 cases localized in every run (see Section~\ref{sec:results-rq3}). The human reference is also not fixed: the annotators agreed with each other less on the introducing write ($\kappa = 0.58$) than on provenance ($\kappa = 0.78$) (see Table~\ref{tab:rq2-summary}(c)), and agreement with R1 varied across task-disjoint subsets (central 95\% of values: 74.42\%--96.88\% for provenance; see Table~\ref{tab:rq2-summary}(b)). High repeatability should therefore not be read as agreement with human judgement, nor the reverse; we suggest reporting both, each for the outputs it covers.

\textbf{Preserving intermediate artefact states matters for agent framework design (RQ3).} Final-only traces had the lowest agreement in all three comparisons (see Figure~\ref{fig:rq3-main}(b)): 4/7 on the introducing write against R1, and 2/4 and 1/4 on the exact origin turn under repeated scoring and between judges. A likely reason is that, without intermediate snapshots, the write that introduced the vulnerable fragment cannot be identified from the artefact history, and the SAGE rule uses this write to bound the origin candidates. Missing states may thus weaken both the artefact-derived outputs and the anchor for the origin turn. Per-write traces, which preserve every intermediate state, still had lower agreement than single-write traces in all three comparisons, so part of the difference may reflect the number of writes in a trace and not only missing states. The final-only group is small ($n = 7$), but the direction was the same in comparisons that rest on different sources of evidence. For framework design, this suggests recording a snapshot after each write and keeping the boundaries of individual writes, so that reasoning turns can be related to code changes during an audit.

\textbf{Implications for future evaluation of agentic repair systems.} Taken together, the results suggest four practices: (i) security evaluation should not stop at the delivered patch, since test passage and final-code checks show whether a repair is insecure, whereas a trace-based analysis adds where the divergence first became observable (see Table~\ref{tab:rq1-overview} and Figure~\ref{fig:rq1-main}); (ii) the origin type and the exact origin turn should each be reported with its own consistency estimate (see Table~\ref{tab:rq2-summary}(a)); (iii) consistency and human agreement should both be reported and kept apart, since they cover different outputs (see Tables~\ref{tab:rq2-summary} and~\ref{tab:rq3-groups}); and (iv) results should be reported by reconstruction condition and by subgroup, with small groups given as counts, since aggregate agreement can hide differences in the available evidence (see Figure~\ref{fig:rq3-main} and Table~\ref{tab:rq3-groups}). Later turns suggest a further direction. At least one missed detection opportunity was flagged in 59 of the 95 cases (see Table~\ref{tab:rq1-overview}(d)), as in the sorting-repair case in which the review did not address the remaining unsafe configuration (see Section~\ref{sec:results-rq1}). This is a pattern observed in our corpus and not a general conclusion about reviewer agents; whether an explicit security requirement check in review turns reduces silent failures, and at what cost, remains to be evaluated in future work. SAGE's origin assignments can thus inform hypotheses about where to intervene, but whether such interventions prevent silent failures requires controlled intervention studies beyond the present observational analysis.

\subsection{Threats to Validity}
\label{sec:threats}

\textbf{Construct validity.} The origin turn is an operational construct: the earliest observable, trace-attributable divergence under the SAGE rule. It does not establish that the selected turn caused the vulnerability or that intervening there would have prevented it, and incomplete records may conceal earlier divergence. A zero score means that no relevant reasoning was recorded, not that the agent had none, and a missed detection opportunity marks a low-scoring later turn, not a turn at which detection would have succeeded. Human annotation covers only the artefact-derived outputs; origin type and turn are assessed through repeated scoring and cross-judge comparison, which measure consistency rather than correctness. As the annotators agreed less on the introducing write ($\kappa = 0.58$) than on provenance ($\kappa = 0.78$), we report agreement with each annotator separately and treat consensus-subset results as conditional.

\textbf{Internal validity.} LLM-judge scoring may introduce run-to-run variation, dependence on the judge model, or hindsight bias. We addressed these risks in five ways: prefix-only scoring; a verbatim quotation for every non-zero score; three scoring runs by the primary judge on a 60-case subset; a second judge from a different model family on the same subset; and comparison of the artefact-derived outputs with two annotators who did not see the SAGE outputs (see Table~\ref{tab:rq2-summary}). These measures reduce but do not remove the threat: the two judges jointly localized 50 of the 60 cases with 78.00\% exact-turn agreement, and biases shared by both judges, inherited from the rubric, or arising from family overlap with the evaluated base models (the primary judge with GPT-5.5 and the second judge with Claude Opus 4.8) cannot be excluded. The annotators used the same reconstructed histories as SAGE, so shared reconstruction errors may inflate agreement, and R1 took part in corpus construction and rule design; we therefore report agreement with R2 separately. As confirmed cases concentrate in few tasks, we assessed sensitivity to task composition with task-disjoint resampling (see Section~\ref{sec:results-rq2}).

\textbf{External validity.} The evaluation covers 95 confirmed silent failures from Python repair tasks in two security-focused benchmarks, six agent frameworks, and six base models, and may not generalize to other languages, task types, or deployment settings. The corpus is bounded by candidate screening: vulnerable patches flagged by neither static analysis nor exploitability-oriented screening do not enter manual review. The confirmed-failure subsets differ in task, vulnerability, and trace composition, so cross-framework and cross-model differences are descriptive and do not estimate framework- or model-level effects. Estimates for groups with fewer than ten cases (see Table~\ref{tab:rq3-groups}; Figure \ref{fig:rq3-main}), including final-only traces, are particularly uncertain, which also qualifies the artefact-preservation implication in Section~\ref{sec:implications}.

\section{Conclusion}
\label{sec:conclusion}

This paper presented SAGE, a trace-based method for localizing the origin of a confirmed silent failure in agentic vulnerability repair. The results show that a divergence from the task's security intent can become observable in the agent's plans and explanations before the vulnerable code is written, and that for inherited vulnerabilities it can be localized even when no code change marks it. Final-patch checks therefore show that a repair is insecure, but not where the repair process first diverged. The results also show that the origin type is reproduced more consistently than the exact origin turn, and that judge consistency and agreement with human annotators cover different outputs, so both should be reported. Agreement was lowest for traces that kept only the final file, which suggests that agent frameworks should preserve intermediate code states. Future work will extend the analysis to other programming languages and task types, and will examine through controlled studies whether safeguards placed at the identified origins prevent silent failures.
\section*{Data Availability}
The dataset is available at~ \citep{anonymous_authors_2026_23108850}.

\bibliographystyle{ACM-Reference-Format}
\bibliography{references}

\end{document}